\documentstyle[12pt]{article}
\begin{document}
\begin{center}
\textbf{SELF-SCREENING  HAWKING  ATMOSPHERE  IN  THE PRESENCE  OF  A  BULK  VISCOSITY}\\
\bigskip
\bigskip
I. Brevik\footnote{E-mail: iver.h.brevik@mtf.ntnu.no}\\
\bigskip
\bigskip
Division of Applied Mechanics, Norwegian University of Science and Technology,\\
N-7491 Trondheim, Norway\\
\bigskip
Revised version, February 2000

\bigskip


\end{center}

\bigskip

\begin{abstract}
The recent theory of 't Hooft [ Nucl. Phys. Suppl. {\bf 68}, 174 (1998)] models the black hole as a system endowed with an envelope of matter that obeys an equation of state in the form $ p=(\gamma -1)\rho$, and acts as a source in Einstein's equations. The present paper generalizes the 't Hooft theory so as to take into account a bulk viscosity $\zeta$ in the fluid. It is shown that even a slight positive value of $\zeta$ will suffice to yield complete agreement with the Hawking formula for the entropy of the black hole, if the value of the constant $\gamma$ takes a value that is slightly less than $4/3$.  The value $\gamma=4/3$ corresponds to a radiation fluid.
\end{abstract} 

\section{Introduction}

The recent paper of 't Hooft \cite{hooft98} is an interesting extension of the usual theories of Hawking radiation \cite{hawking75}. The 't Hooft model is static. The Hawking particles emitted by a black hole are treated as an envelope of matter (a static fluid) that obeys an equation of state, and acts as a source in Einstein's equations. The equation of state is conventionally written as $p=(\gamma -1)\rho$, with $\gamma$ a constant lying between 1 and 2. Here $\gamma =1$ corresponds to a pressure-free fluid, $\gamma =4/3$ corresponds to a radiation fluid, and $\gamma = 2$ yields the Zel'dovich fluid in which the velocity of sound equals the velocity of light. Another related work that ought to be referred to in the present context, and which is actually prior of that of 't Hooft,  is the one of Zurek and Page \cite{zurek84}.

There exists evidently a very natural generalization of 't Hooft's theory, namely to take into account the {\it time dependence} of the Hawking evaporation process. A black hole emitting particles necessarily has to lose mass; accordingly the static fluid model envisaged by 't Hooft can only be an approximation. In the present paper we will focus attention on one particular aspect of the time-dependent generalization of 't Hooft's theory, namely the influence from a possible {\it viscosity} of the self-screening atmosphere. Although we do not know at present how large the viscosity of the atmospheric fluid actually is, it seems nevertheless worthwhile to examine the physical influence from this factor. From a general hydrodynamic point of view it is actually somewhat surprising to note how little attention is usually being paid to the viscosity concept in relativistic or cosmological fluids; after all, we know that viscosity often plays an important role in fluid mechanics. In the cosmological context, there are a few studies of the influence from viscosity; we may mention the extensive treatments of Weinberg \cite{weinberg71}, and of Gr{\o}n \cite{gron90}, and there are a few others.

As we will see, the incorporation of a bulk viscosity $\zeta$ leads to the following attractive physical property of the theory: the Hawking formula for the entropy of the black hole, which was rather unnatural to reproduce in the static 't Hooft formalism, can be incorporated in the present formalism in a straightforward way even if the influence from viscosity is very slight. If $\zeta$ is chosen to have some small, constant, positive  value, and if the  quantity $\dot \beta e^{-\alpha}$ (see Eq.(38) below) is roughly taken to be a constant, then $\gamma$ turns out to be a constant, slightly less than 4/3. That is, $\gamma$ is slightly less than the value corresponding to a radiation fluid.

\section{Einstein's equations}

Let us thus assume the existence of a spherically symmetric atmospheric fluid, whose properties vary very slowly with time, in the region around the black hole. We recall the characteristic properties of the 't Hooft model: the fluid is most dense around the would-be horizon at $r=2M$, but is otherwise present everywhere, on the outside of the horizon as well as on the inside of it, except at the origin where there resides a negative mass. (This negative mass is the price paid for the existence of the atmospheric blanket.) We write the line element in the form
\begin{equation}
ds^2=-e^{2\alpha(r,t)}dt^2+e^{2\beta(r,t)}dr^2+r^2(d\theta^2+\sin^2 \theta d\varphi^2).
\end{equation}
\label{1}
Einstein's equations are, with $G=1$,
\begin{equation}
G_{\mu\nu}\equiv R_{\mu\nu}-\frac{1}{2}g_{\mu\nu} R=8\pi T_{\mu\nu}.
\end{equation}
\label{2}
Consider first the Einstein tensor $G_{\hat\mu \hat\nu}$ in an orthonormal basis.
(Reference \cite{lightman75}, for instance, gives useful expressions for this tensor.) We shall need the expressions for the mixed components:
\begin{equation}
G_{\hat 0}^{\hat 0}=-\frac{1}{r^2}+\frac{e^{-2\beta}}{r^2}(1-2\beta' r),
\end{equation}
\label{3}
\begin{equation}
G_{\hat r}^{\hat 0}=-\frac{2\dot \beta}{r}e^{-(\alpha +\beta)},
\end{equation}
\label{4}
\begin{equation}
G_{\hat \theta}^{\hat 0}=G_{\hat \varphi}^{\hat 0}=G_{\hat \theta}^{\hat r}=G_{\hat \varphi}^{\hat r}=0,
\end{equation}
\label{5}
\begin{equation}
G_{\hat r}^{\hat r}=-\frac{1}{r^2}+\frac{e^{-2\beta}}{r^2}(1+2\alpha' r),
\end{equation}
\label{6}
\begin{equation}
G_{\hat \theta}^{\hat \theta}=G_{\hat \varphi}^{\hat \varphi}=-e^{-2\alpha}(\ddot \beta +\dot \beta ^{2}-
\dot \alpha \dot \beta)+e^{-2\beta}(\alpha ''+\alpha'^{2}-\alpha' \beta' +\frac{\alpha'-\beta'}{r}),
\end{equation}
\label{7}
primes and dots meaning derivatives with respect to $r$ and $t$. Our basic assumption that the fluid, and consequently also the metric, change "slowly", comes into play only in Eqs. (4) and (7). The other expressions are exactly as in the static case.

In accordance with our slowness assumption we can neglect the first and second order time derivatives of $\beta$ in comparison to unity. Moreover, we shall assume that $\dot \beta$, when multiplied with the bulk viscosity $\zeta$, stays finite. Thus, the time scale for our "slow" time variations is characterized by
\begin{equation}
|\dot \beta| \ll 1,~~~|\ddot \beta| \ll 1,~~~\zeta \dot \beta = {\rm finite}.
\end{equation}
\label{8}
We can accordingly neglect $\ddot{\beta},~\dot{\beta}^2$, and $\dot \alpha \dot\beta$ In Eq. (7). This means that $G_{\hat \theta}^{\hat \theta}=G_{\hat \varphi}^{\hat \varphi}$ reduce to the same form as for a static fluid, whereas $G_{\hat r}^{\hat 0}=0$. 

It is of interest to relate the assumptions of Eq. (8) to the rate of energy dissipation in the fluid. From classical fluid dynamics it is known that the rate of energy dissipation per unit volume, caused by bulk viscosity, is $\dot \epsilon = \zeta (\bf{\nabla \cdot u})^2 $ (cf. Sec. 79 in \cite{landau87}). Thus, in our case the analogous expression per unit proper volume becomes (cf. Eq. (15) below):
\begin{equation}
\dot \epsilon = \zeta \theta^2 = \zeta \dot{\beta}^2 e^{-2\alpha}.
\end{equation}
\label{9}
From this expression we see, however, that $\dot{\epsilon}$ is a negligible quantity in our approximation, because of the extra factor $\dot{\beta}$ appearing in the last term on the right.

Consider next the energy-momentum tensor $T_{\mu\nu}$. As mentioned, the fluid acts as a source in (2). Let $U^\mu = (U^0,U^{i})$ be the four-velocity of the fluid. We ignore spatial derivatives, but keep time derivatives, of $U^\mu$. In the static coordinate system we work in, we can set $U^0=(-g_{00})^{-1/2}=e^{-\alpha}$, $U_0=-e^{\alpha}$, $U^{i}=0$, since the fluid is practically at rest. We introduce the scalar expansion $\theta$, the projection tensor $h_{\mu\nu}$, and the shear tensor $\sigma_{\mu\nu}$:
\begin{equation}
\theta=U_{;\mu}^\mu ~,~~~~ h_{\mu\nu}=g_{\mu\nu}+U_{\mu}U_{\nu},
\end{equation}
\label{10}
\begin{equation}
\sigma_{\mu\nu}=\frac{1}{2}(U_{\mu;\alpha}h_{\nu}^{\alpha}+U_{\nu;\alpha}h_{\mu}^{\alpha})-\frac{1}{3}\theta h_{\mu\nu}
\end{equation}
\label{11}
(cf, for instance, \cite{lightman75} or \cite{brevik94}). Then, if $\eta$ is the shear viscosity and $\zeta$ the bulk viscosity, we can write $T_{\mu\nu}$ as
\begin{equation}
T_{\mu\nu}=\rho U_{\mu}U_{\nu}+(p-\zeta \theta)h_{\mu\nu}-2\eta \sigma_{\mu\nu} ,
\end{equation}
\label{12}
assuming constant temperature in the fluid. There are thus {\it two} viscosity coefficients. Usually, in cosmological applications one exploits the assumed isotropy of the fluid to omit the shear viscosity term. It ought to be stressed, however, that this point is more delicate than what is often recognized. The reason is that the shear viscosity under usual cosmological conditions is very much greater than the bulk viscosity. As an example, we mention that that in the universe, after termination of the plasma era at the time of recombination ($T \simeq 4000$ K) one has \cite{brevik94}
\begin{equation}
\eta_{recomb} \simeq 6.8 \times 10^9~g\,cm^{-1}\,s^{-1} ,~~~~\zeta_{recomb} \simeq 2.6\times 10^{-3}~g\,cm^{-1}\,s^{-1}.
\end{equation}
\label{13}
The shear viscosity at this instant thus outweighs the bulk viscosity by about 12 orders of magnitude. Accordingly, even a very slight anisotropy in the fluid may easily compensate for the influence from the bulk viscosity.

Notwithstanding this remark, we shall in the following follow cosmological practice and include the bulk viscosity only, thus assuming strict isotropy, in order to keep the formalism as simple as possible. Observing the relations
\[ \Gamma_{00}^0=\dot\alpha,~~~~~\Gamma_{rr}^0 = \dot\beta e^{2(\beta-\alpha)},\]
\begin{equation}
U_{0;0}=0,~~~~~U_{r;r}=\dot \beta e^{2\beta-\alpha},
\end{equation}
\label{14}
we find for the scalar expansion
\begin{equation}
\theta= U_{;0}^0+U_{;r}^r=\dot\beta e^{-\alpha}.
\end{equation}
\label{15}
We now define the effective pressure $\tilde p$:
\begin{equation}
\tilde p \equiv p-\zeta \theta= p-\zeta\dot\beta e^{-\alpha},
\end{equation}
\label{16}
whereby we can write the energy-momentum tensor as
\begin{equation}
T_{\mu\nu}=\rho U_{\mu} U_{\nu}+\tilde p h_{\mu\nu}.
\end{equation}
\label{17}
Einstein's "energy" equation in the orthonormal frame, $G_{\hat 0}^{\hat 0}=8\pi T_{\hat 0}^{\hat 0}$, leads to
\begin{equation}
1-e^{-2\beta}(1-2\beta' r)=8\pi \rho r^2,
\end{equation}
\label{18}
which is the same equation as if the fluid were non-viscous. The analogous "pressure" equation, $G_{\hat r}^{\hat r}=8\pi T_{\hat r}^{\hat r}$, leads to
\begin{equation}
1-e^{-2\beta}(1+2\alpha' r)=-8\pi\tilde p r^2,
\end{equation}
\label{19}
in which the influence from viscosity is present explicitly on the right hand side. Finally, the equation $G_{\hat \theta}^{\hat \theta}=8\pi T_{\hat \theta}^{\hat \theta}$ leads to
\begin{equation}
e^{-2\beta}(\alpha''+\alpha'^2-\alpha' \beta'+\frac{\alpha'-\beta'}{r})=8\pi \tilde p,
\end{equation}
\label{20}
when account is taken of the two first members of Eq. (8). These equations are the same as in the time-independent case, only with the replacement $p\rightarrow \tilde p$.

\section{Discussion on the solutions of the equations}

We insert Eq.(17) into the energy-momentum conservation equation ${T^{\mu\nu}}_{;\nu}=0$ and multiply with $h_{\alpha\mu}$. Since $h_{\alpha\mu}U^{\mu}=0$ we obtain
\begin{equation}
h_{\mu}^{\nu} \tilde p_{,\nu}=-(\rho+\tilde p) U_{\mu;\nu} U^{\nu}.
\end{equation}
\label{21}
We put $\mu = r$, and observe $U_{r;0}=-\Gamma_{r0}^0 U_0=\alpha' e^{\alpha}$ to obtain
\begin{equation}
\tilde p'=-(\rho +\tilde p)\alpha'.
\end{equation}
\label{22}
This is the Tolman-Oppenheimer-Volkoff equation, with the replacement $p\rightarrow \tilde p$. Making use of this equation, we see that Eq. (20) is a consequence of Eqs. (18) and (19). That is, we can henceforth consider Eqs. (18) and (19) to be the basic governing equations. This is the same kind of behaviour as in the static case.

Let us for a moment ignore viscosity.  We then see that the conventional state equation for cosmic fluids,
\begin{equation}
p=(\gamma -1)\rho,
\end{equation}
\label{23}
when inserted into (22), yields after integration
\begin{equation}
\rho e^{\frac{\alpha\gamma}{\gamma-1}}=const.
\end{equation}
\label{24}
In particular, if $\gamma=4/3$ as  in the case of massless particles, we obtain $\rho e^{4\alpha}=const.$ This agrees with Eq. (2.6) in \cite{hooft98} (the notation is different).

Now reinstating viscosity, we have to assume a relation between $\tilde p$ and $\rho$ in order to solve (22). We adopt henceforth the ansatz that most closely lies at hand, namely to set the deviation in pressure, $\zeta \dot \beta e^{-\alpha}$, proportional to the pressure itself. The proportionality constant will be called $\xi$. Thus, we put
\begin{equation}
\tilde p = (1-\xi)p.
\end{equation}
\label{25}
In turn, this means that we can write a natural equation of state, analogous to (23), also for the viscous fluid, but in terms of "tilde" variables $\tilde p$ and $\tilde \gamma$ instead of the usual $p$ and $\gamma$:
\begin{equation}
\tilde p = (\tilde \gamma -1)\rho .
\end{equation}
\label{26}
Here
\begin{equation}
\tilde \gamma = \gamma-\xi (\gamma-1).
\end{equation}
\label{27}
It is to be noted that our adoption of the linear (barytropic) equation of state in the form (23) or (26) is intended to be {\it compatible} with the presence of the bulk viscosity. The formal similarity with the equation of state for an ideal (nonviscous) fluid is not to be taken as if we mean identifying the fluid with an {\it isentropic} fluid. As is generally known, the presence of any of the two possible viscosity coefficients means that there is a continous production of entropy in the fluid; cf. Eq. (49.6) in \cite{landau87}. The state equation $p=(\gamma -1)\rho$ has frequently been made use of also in earlier works on viscous cosmology; cf., for instance, \cite{gron90, brevik94, burd94}.

Instead of (24) we now obtain, by integration of (22),
\begin{equation}
\rho e^{\frac{\alpha \tilde \gamma}{\tilde \gamma -1}}=const. \equiv \frac{C}{8\pi}.
\end{equation}
\label{28}
So far, no restriction has been made on the magnitude of $\xi$. Let us from now on assume, what seems most natural, that the influence from the bulk viscosity is so small that we can take $\xi \ll 1$. It becomes then possible in principle to solve the governing equations (18) and (19), choosing reasonable values for the input parameters. One obvious possibility would be to put $\gamma=4/3$, whereby $\tilde \gamma= 4/3-\xi/3$. Let us consider this possibility first. Then
\begin{equation}
\rho e^{4\alpha}(1+3\xi \alpha)=\frac{C}{8\pi},
\end{equation}
\label{29}
to first order in $\xi$. We can scale Einstein's equations by introducing new variables $X$ and $Y$:
\begin{equation}
X=\frac{e^{2\alpha}}{r} ,~~~~~Y=e^{2\beta}.
\end{equation}
\label{30}
Equations (19) and (18) can now be written, to order $\xi$, as
\begin{equation}
\frac{rX'}{X}=Y-2+\frac{CY}{3X^2}\left[ 1-\xi(1+\frac{3}{2}\ln (rX)) \right],
\end{equation}
\label{33}
\begin{equation}
\frac{rY'}{Y}=1-Y+\frac{CY}{X^2}\left[ 1-\frac{3}{2}\xi \ln (rX) \right].
\end{equation}
\label{32}
These equations agree with Eqs. (2.9) and (2.10) in \cite{hooft98} in the non-viscous case, $\xi=0$.
 Let the solutions in the last-mentioned case be denoted by $X_0$ and $Y_0$.  As $r$ can now be eliminated from the equations, it is convenient to establish a nonlinear differential equation between $X_0$ and $Y_0$:
\begin{equation}
\frac{dY_0}{dX_0}=\frac{Y_0(1-Y_0+\frac{CY_0}{X_0^2})}{X_0(Y_0-2+\frac{CY_0}{3X_0^2})}.
\end{equation}
\label{33}
This equation can be integrated in the inward direction, starting from large values of $r$ for which
\begin{equation}
X_0=\frac{1}{r}(1-\frac{2M}{r}), ~~~~~Y_0=(1-\frac{2M}{r})^{^{-1}}.
\end{equation}
\label{34}
The solution was given in \cite{hooft98}, and will not be further considered here. As mentioned, it leads to a black hole of negative mass at the origin. Let us instead consider the viscous case: then, the situation becomes mathematically more complex as we cannot any longer eliminate the variable $r$ from the governing equations (31) and (32). It is possible, however, to work out a first order perturbative expansion in $\xi$, by writing
\begin{equation}
X=X_0(1+\xi X_1), ~~~~~Y=Y_0(1+\xi Y_1),
\end{equation}
\label{35}
with $X_1$ and $Y_1$ being zeroth order quantities. From (31) and (32) we derive the first order "pressure" equation
\begin{equation}
rX_1'=Y_0Y_1+\frac{CY_0}{3X_0^2}\left[ Y_1-2X_1 -1-\frac{3}{2}\ln (rX_0) \right],
\end{equation}
\label{36}
and the first order "energy" equation
\begin{equation}
rY_1'=-Y_0 Y_1+\frac{CY_0}{X_0 ^2}\left[ Y_1-2X_1-\frac{3}{2}\ln (rX_0) \right].
\end{equation}
\label{37}
Once $X_0$ and $Y_0$ are known as functions of $r$, we can in principle calculate $X_1$ and $Y_1$ by integrating these two equations, again starting with large values of $r$ and integrating in the inward direction. As initial conditions for large $r$ we may set $X_1=Y_1=0$. If the value of the parameter $\xi$ is known (in practice it has to be chosen), we can thus finally find the scaled metric from Eq. (35).

However, we shall not work out the solution in this case which, as mentioned, was  based upon the choice $\gamma =4/3$, in detail. What seems to be of greater  physical interest, is the determination of the value of $\xi$ for which complete match can be obtained with Hawking's formula for the {\it entropy of the black hole}. In Planck units, the entropy per unit surface area is according to Hawking equal to $1/4$. As stated by 't Hooft \cite{hooft98}, this case requires the constant $\gamma$ for a non-viscous fluid to have the value 2. In other words, the fluid has to be of the Zel'dovich type. As mentioned in \cite{hooft98}, it is difficult to imagine ordinary matter with such a high $\gamma$ value.

So, the central point of our paper becomes the following: for which value of $\xi$, and thereby also the viscosity, do we get perfect match with the mentioned Hawking formula? Actually, we do not need to make any further calculation to answer this question, since Eqs. (18) and (19) differ formally from the corresponding non-viscous equations only in the replacement $p\rightarrow \tilde p$. Consequently, if
\begin{equation}
\tilde p=\frac{1}{3}\rho 
\end{equation}
\label{38}
in Eq. (19) we can carry out the same analysis as in \cite{hooft98} and obtain exactly the Hawking entropy. Since then $\tilde \gamma = 4/3$, we obtain from Eq. (27) the relationship
\begin{equation}
\xi \equiv \zeta \dot \beta e^{-\alpha}=\frac{\gamma-4/3}{\gamma-1}.
\end{equation}
\label{39}
Since $\xi$ is assumed small, we can here replace $\alpha$ by $\alpha_0$. Thus, in order to obtain complete match with the Hawking entropy the only condition is that Eq. (39) has to be satisfied. The values of $\xi$ and $\gamma$ by themselves do not have to be fixed. As a working hypothesis we may roughly assume that the quantity $ \dot \beta e^{-\alpha_0}$ is constant; then the bulk viscosity $\zeta$, as well as the parameter $\gamma$, turn out to be constants. The constancy of $\gamma$ should be expected, in view of the equation of state for the fluid, adopted as it is in the form (23).

One may ask: is the expression (39) physically reasonable? The answer turns out to be in the affirmative, due to the following reason: in a realistic fluid, we expext that the value of $\gamma$ in the state equation $p=(\gamma -1)\rho $ lies somewhere between 1 (pressure-less fluid) and 4/3 (radiation fluid). This means that the right hand side of (39) is negative. Moreover, due to the emission of matter from the black hole we must have $\dot M < 0$, and this corresponds to $\dot \beta < 0$ for a given value of $r$. [ At large values of $r$, this follows from the last member of Eq. (34). Also in the main region of the atmospheric blanket, centered around $r=2M$, we have $\dot \beta <0$; cf. Fig. 2 in \cite{hooft98}. The only exception may be in the vicinity of the origin.] Consequently, Eq. (39) yields $\zeta >0$.  The bulk viscosity turns out to be {\it positive}, as it should be according to ordinary thermodynamics \cite{landau87}. The positiveness of the viscosity coefficients comes from the general thermodynamic property that the entropy change for an irreversible process in a closed physical system is always positive.

Consider finally the question how to determine the location of the horizon of the time-dependent black hole. We assume that the static (i. e., the non-viscous) problem has been solved, so that $X_0$ and $Y_0$ are known functions of $r$. Also, the constant $C$, appearing in Eq. (28), is then known. The "horizon"  in the presence of an envelope of matter is not a singular boundary on which the metric diverges, but is naturally defined as the surface where the function $e^{2\beta}$ has a maximum (cf. the metric in Eq. (1)). That is, the horizon corresponds in the static case to the equation $Y_0'=0$. It is given explicitly as a dashed line in Fig. 2 in \cite{hooft98}. Mathematically, we can write the condition for the static horizon as
\begin{equation}
1-Y_0+\frac{CY_0}{X_0^2}=0
\end{equation}
\label{40}
(cf. Eq. (2.10) in \cite{hooft98}, or Eq. (32) above with $\xi=0$). Moving on to the time-dependent, viscous case, the horizon is analogously determined by the equation $Y'=0$, or
\begin{equation}
Y_0'+\xi(Y_0Y_1)'=0;
\end{equation}
\label{41}
cf. Eq. (35). The position of the horizon is seen to be slightly displaced relative to the static case because of the constant small factor $\xi=\zeta \dot\beta e^{-\alpha_0}$. For practical purposes it is more convenient to express the condition (41) in such a form that the functions $X_0,~Y_0,~X_1,~Y_1$ occur, but not their derivatives: taking the logarithmic derivative of Eq. (35), $d \ln Y/(d\ln r) \simeq d\ln Y_0/(d\ln r)+\xi dY_1/(d\ln r)$ we get, using Eq. (37), 
\begin{equation}
1-Y_0+\frac{CY_0}{X_0^2}
-\xi Y_0\left[ Y_1-\frac{C}{X_0^2}\left(Y_1-2X_1-\frac{3}{2}\ln (rX_0)\right) \right]=0. 
\end{equation}
\label{42}
If the set of equations (36) and (37) has been solved, so that $X_1$ and $Y_1$ are known functions of $r$, Eq. (42) thus fixes the position of the horizon. In practice, numerical work becomes necessary.

\section{Concluding remarks}

Of course, our theory presented above is somewhat speculative. Already the static 't Hooft model \cite{hooft98} is to some extent speculative. The 't Hooft model is however an interesting alternative to the standard picture of a black hole, since it explicitly takes into account the influence upon the metric from the atmospheric blanket consisting of the emitted Hawking particles. Although it is not evident in advance that the introduction of viscosity coefficients in the blanket is needed in the model, this idea does not seem unreasonable to us in view of the general importance of viscosities in ordinary hydrodynamics.

One important observation in our paper is that Einstein's "pressure" equation (19) takes the same form as the corresponding equation in the non-viscous case, only with the substitution  $p \rightarrow \tilde p$, where $\tilde p$ is defined by Eq.(16). This makes it possible to use 't Hooft's results and write down the formula (39), giving agreement with Hawking's entropy, directly. Thus only a slight bulk viscosity is sufficient to give the Hawking entropy, if the value of $\gamma$ is adjusted accordingly, according to Eq. (39). A small viscosity corresponds to a value of $\gamma$ being slightly less than 4/3, meaning that the velocity of sound becomes slightly less than $c/\sqrt {3}$. This seems physically reasonable. Moreover, we obtain a {\it positive} bulk viscosity, which is in agreement with ordinary thermodynamics.

Of course, one may wonder why we find it so desirable to maintain the connection with Hawking's formula for the entropy of a black hole. We do not enter into a detailed discussion of this point, but restrict ourselves to mentioning that Hawking's formula seems to be the outcome of investigations along different routes, and therefore ought to be regarded with some confidence in the difficult field of gravitational thermodynamics.

Finally, it is to be noted that quantum corrections to the thermodynamics of the 't Hooft black hole model have recently been calculated by Nojiri and Odintsov \cite{nojiri99}. The inclusion of quantum corrections does not change the qualitative properties of this model. For example, the area law is found to be the same as without quantum corrections. 

{\bf Acknowledgement.}
I wish to thank Professor Sergei Odintsov for valuable information about this problem.

\end{document}